\documentclass{PoS}

\newcommand{\fermi}{\textit{Fermi}-LAT}
\newcommand{\hess}{H.E.S.S.}

\newcommand{\E}[1]{\times 10^{#1}}
\newcommand{\g}{\ensuremath{\gamma}}

\title{Blazar variability -- expect the unexpected}

\ShortTitle{Blazar variability}

\author{\speaker{Michael Zacharias}\\
        Ruhr Astroparticle and Plasma Physics Center (RAPP Center), Insitut f\"ur theoretische Physik IV, Ruhr-Universit\"at Bochum, D-44780 Bochum, Germany\\
        Centre for Space Research, North-West University, Potchefstroom, 2520, South Africa \\
        E-mail: \email{mzacharias.phys@gmail.com}, \email{mz@tp4.rub.de}
        }

\abstract{Despite significant progress in both data taking and the development of theory, blazars keep offering surprises. While many frequency bands are now covered by long-term monitoring efforts on at least a few selected sources, blazars remain unpredictable and the same source may exhibit vastly different behaviors. Here, an overview is given about four selected sources, namely 3C\,279, PKS\,2155-304, PKS\,1510-089 and CTA\,102. All of them exhibited, indeed, highly unexpected events.
}

\FullConference{High Energy Astrophysics in Southern Africa - HEASA2018\\
		1-3 August, 2018\\
		Parys, Free State, South Africa}

\begin{document}
\section{Introduction}
Variability of blazars, the relativistically beamed version of active galaxies \cite{br74}, is observed across all frequencies and time scales. Flares as short as minutes, as well as high states lasting for months to years are a common feature. Apart from the duration of flares, the specific form of the lightcurve can be different for each flare (symmetric, fast rise and slow decay, vice versa, etc). Interestingly, flares within an individual source are different from case to case and it is difficult to connect a given flare with a specific mechanism. 

Nonetheless, blazars share common features. Most notably, the characteristic double-humped spectral energy distribution (SED). The low-energy component is attributed to synchrotron emission of relativistic electrons, while the high-energy component is either due to inverse Compton (IC) emission of the same electron population or processes involving relativistic protons. Seed photons for the former process are the internal synchrotron photons (resulting in synchrotron-self Compton (SSC) emission \cite{mgc92}) or external photon fields, such as the accretion disk \cite{ds93}, the broad-line region (BLR) \cite{sbr94}, the dusty torus (DT) \cite{bea00}, the cosmic microwave background \cite{bdf08} and others, if available. Hadronic processes involve proton synchrotron emission, proton-$\gamma$ interactions producing pions, muons, and electrons\footnote{The term electrons also includes positrons for convenience.}, as well as Bethe-Heitler pair cascades \cite{mp01,bea13}. Different combinations of these processes are usually considered within a one-zone emission model, where a small zone within the large-scale jet is responsible for the main radiative output. The short variability time scales, and specific large-scale jet models \cite{pc13,lea18}, suggest that the emission is, indeed, dominated by specific regions within the jet\footnote{Extended jet emission, as in the blazar AP Librae \cite{kwt13,zw16}, are ignored here, as no significant variability is expected from extended jets.}.

Blazars are classified according to the width of their optical emission lines into flat spectrum radio quasars (FSRQs), which exhibit broad lines ($EW>5\,$\AA), and BL Lac objects, which show narrow or no lines. BL Lac objects are further classified according to the peak frequency of their synchrotron component into low-frequency peaked BL Lac objects (LBL, $\nu_{\rm peak}<10^{14}\,$Hz), intermediate-frequency peaked BL Lac objects (IBL, $10^{14}\,$Hz$\,<\nu_{\rm peak}<10^{15}\,$Hz) and high-frequency peaked BL Lac objects (HBL, $\nu_{\rm peak}>10^{15}\,$Hz). 

In this article, a selected number of particularly interesting flares are summarized along with observational features and potential explanations. These flares are among the most extreme cases of blazar outbursts, and by no means display ``normal'' blazar behavior. Nonetheless, it is these cases where our understanding of jets is pushed to the limits.

%
%
\section{3C\,279}
3C\,279 is an FSRQ at a redshift of $z = 0.536$ \cite{br65,msdcm96}. It was first detected at high-energy (HE, $E>100\,$MeV) $\gamma$-rays with EGRET \cite{hea99} and has been established as a highly variable source in this energy range with \fermi\ \cite{aFea15}. It produced its first big surprise by being the first FSRQ to be detected at very-high-energy (VHE, $E>100\,$GeV) \g-rays with MAGIC in 2006 \cite{Mea08}, as the bright nuclear photon field of the BLR was assumed to attenuate all VHE photons. The VHE \g-ray emission has strong implications for the modeling of the flare, and simple leptonic one-zone models failed to reproduce the spectrum \cite{brm09}. Either leptonic multi-zone or hadronic models were required to account for the observations.

The advent of the \fermi\ instrument in August 2008 resulted in a continuous coverage of 3C\,279 at HE \g-rays. Contemporaneous multiwavelength (MWL) observations revealed a large variety of flaring characteristics. The HE \g-ray spectrum derived for various flares shows all kinds of behaviors from simple flux increases to flux increases with softer spectra, and even flux increases with significantly harder spectra. The latter can lead to photon indices as hard as $1.7$ \cite{Hea15}. These hard spectra might explain the potential for VHE \g-ray detections in 3C\,279.

The brightest detected flare at HE \g-rays took place in June 2015. Integrated fluxes reached values exceeding $3\E{-5}\,$ph/cm$^2$/s in a $3\,$hr time bin, which is at least a factor $30$ above quiescent levels and made 3C\,279 briefly the brightest \g-ray source in the sky. The flare resulted in various target-of-opportunity (ToO) observations, including Swift and INTEGRAL. The resulting time-averaged spectra were interpreted using leptonic and hadronic steady-state one-zone models. While both models can fit the data, the leptonic model suffers from low, sub-equipartition magnetic fields, and the hadronic model from a large jet power \cite{bea16}. The high \g-ray fluxes measured with \fermi\ enabled the detection of intra-orbit variability as short as a few minutes \cite{aFea16}. This is challenging for any model. A possible explanation is moving mirrors \cite{vtc17}, where the reflection of radiation back into the emission region enhances the energy density of the reflected radiation by orders of magnitudes. Strong magnetic fields on the order of kG could account for the fast variability within a hadronic model, as shown by explicit modeling \cite{pea17}. The picture of this flare is further complicated by the detection of VHE \g-rays with \hess\ \cite{rea17}. This observation indicates that the emission region must be located outside of the strong BLR photon field, which rules out IC/BLR as the main \g-ray emission process.

3C\,279 has revealed further surprises in subsequent years. In 2017 it underwent a major optical flare \cite{atomAtel17} without a significant counterpart in the \g-ray domain. In early 2018 another big \g-ray flare was recorded with \fermi, and \hess\ observed renewed VHE \g-ray activity -- however, after the HE \g-ray flux had considerably dropped \cite{hessAtel18}. 3C\,279 continued to be highly active in 2018 with two more detections at VHE \g-rays \cite{magicAtel18,hessAtel18-2} and long periods of MWL activity. Apparently, the source is able to produce VHE \g-rays during many flares, but when and how exactly is an open question.
%
%
\section{PKS\,2155-304}
PKS\,2155-304 is an HBL at a moderate redshift of $z=0.116$. It was established as an X-ray source in the late 1970s \cite{gea79,sea79}, and detected at HE \g-rays with EGRET in the 1990s \cite{sv97}. Strong variability was also established by MWL observations at that time with X-ray variability as fast as $1\,$hr \cite{zea02}. During an MWL outburst in 1997, the first detection at VHE \g-rays was reported by the Durham group \cite{cea99}. \hess\ confirmed the VHE \g-ray nature of PKS\,2155-304 in 2002 and 2003 \cite{aHea05a}, and has monitored the source ever since. 

Several MWL campaigns have been done on PKS\,2155-304, however the picture remains inconclusive. In 2003 a campaign lasting several weeks established intranight variability in the VHE domain, but no correlation with any other band could be found \cite{aHea05b}. A similar campaign in 2008 that included for the first time the \fermi\ instrument, revealed evidence for correlated behavior between the VHE and the optical band, but not between the VHE and the X-ray or HE \g-ray band \cite{aHea09a}. As the lightcurves were only mildly variable, PKS\,2155-304 was considered to be in a low state. The data of the 2008 campaign was modeled using a time-dependent two-zone SSC model in order to achieve the different correlations \cite{wrs10}. A similar conclusion was reached by a different study \cite{p14}, which compared a one-zone SSC model with a two-zone SSC model and a lepto-hadronic scenario. As both the two-zone SSC and the lepto-hadronic scenario achieve comparable fits, which are significantly better than the one-zone SSC model, it was concluded that at least two particle populations (two electron populations for the two-zone SSC, or an electron and a proton population in the lepto-hadronic model) are required to explain PKS\,2155-304 in its low state. In line with this statement, a reproduction of the same data set was also achieved using a spine-sheath model \cite{rrm12}, where the sheath contains electrons and protons, while the spine contains electrons and positrons.

Apparently, PKS\,2155-304 is already interesting in its low state, as the simple one-zone SSC model already fails. However, the source is probably most famous for its tremendous VHE \g-ray outburst observed with \hess\ in July 2006. In the early hours of July 28 (MJD~$53944$), the flux reached an average value of more than 10 times of previous quiescent levels, corresponding to more than 7 times the Crab Nebula flux above the same energy threshold ($200\,$GeV). Peak fluxes on minute scales exceeded this average flux by a factor 2 \cite{aHea07}. Two nights later another, and similarly strong, outburst was detected -- this time simultaneous observations were achieved in the VHE \g-rays and X-rays by \hess\ and Chandra, respectively \cite{aHea09b}. The VHE-to-X-ray flux ratio reveals a cubic relation, which cannot be adequately described by the usual SSC process. The second flare was modeled in \cite{aHea12} employing three models. The first one uses a two-zone model, where a larger emitting region is responsible for the X-ray component, while a smaller and faster moving region is responsible for the VHE \g-ray output. The minute-short variability could be attributed to more small zones. In this model, all components are basically independent of each other. A second model used the blob-in-a-jet scenario, where the two zones are no longer independent. The jet is strongly out of equipartition, while the blob is close to equipartition. The jet part of this model includes a large-scale structure that also reproduces the radio portion of the spectrum and incorporates adiabatic losses of a particle distribution moving down the expanding jet. The third model assumes a stratified jet, where particles are injected at the base and their flow down the jet is followed. The lightcurves and spectra have been calculated by integrating over the entire model jet. Most of the variability is accounted for by varying the acceleration strength. It is noted that the absorption within the jet injects electron-positron pairs that increase the particle flux between the base and the ``main'' emission point by an order of magnitude. The pair injection is considered as one of the main drivers of the fast variability in this case. A different scenario for the flare is drawn in \cite{np12}, which considers turbulent flows and a jets-in-jet scenario. In both cases the flaring region contains numerous subregions, which move in random directions with random Lorentz factors. This randomness is able to explain the fast variability.

PKS\,2155-304 is still closely monitored by \hess\ and other instruments. While a flare, similar to that in 2006, has not been seen again, the source's behavior remains full of surprises. An analysis of the low-state reveals log-normal behavior and colored noise \cite{aHea17}, which differs from the red-noise spectrum during the flare implying a break in the power spectrum at time scales of about a day. If this break is a universal feature of the source or stems from the flare itself, is yet to be established.
%
%
\section{PKS\,1510-089}
PKS\,1510-089 is an FSRQ at a redshift $z=0.361$. It was the second established FSRQ at VHE \g-rays \cite{aHea13}. This detection was made with \hess\ during a HE \g-ray and optical flare in 2009. It added a piece to the puzzling picture of the source. The source was very active during the first 8 years of the operation of \fermi, however no clear correlation pattern between any band has emerged yet as the MWL behavior changed from flare to flare, and so did the models used to explain them.

The 2009 flare has been explained in a steady-state scenario by invoking the IC emission of both the BLR and the DT \cite{bea14}. As the Klein-Nishina cut-off prevents any BLR photons to be scattered into the VHE \g-ray regime, the DT photon field is required to explain the detection with the \hess\ experiment. The inevitable absorption of VHE \g-ray photons in the BLR photon field requires that the location of the emission region during this flare was located close to the outer edge of the BLR.

A study \cite{nea12} using \textit{Herschel} data taken in September 2011 revealed that a single zone could not account for the contemporaneous MWL spectrum. Instead, a two-zone model was applied, where the infrared part originated from synchrotron emission by a zone embedded within the DT, while the higher frequency synchrotron emission, as well as the HE \g-ray emission originated from a zone within the BLR.

Three major flares occurred in quick succession in October/November 2011 \cite{sea13}. The detailed HE \g-ray lightcurve reveals that every peak exhibits a different behavior. The first peak is very short, lasting just a few hours. The second peak lasted a few days with an almost symmetric profile. The third peak also lasted a few days, however with a sharp rise, and slow decay. This activity phase has been explained with an expanding emission region outside of the BLR \cite{sea15}. The expansion leads to non-uniformity of the Doppler factor across the jet, which distorts the lightcurve and may explain the different profiles.

A two months long flaring state in 2012 with strong variability in the HE \g-ray and optical R band resulted in ToO observations with MAGIC \cite{aMea14}, which confirmed the \hess\ detection of PKS\,1510-089 as a VHE \g-ray emitter. While the source was highly variable in many bands, no variability was detected in the VHE band (similar to the 2009 flare). Two steady-state modeling attempts have been made for the integrated spectra. The first one is a one-zone model located within the DT to avoid the strong attenuation of \g-rays by the BLR. The second model employs the spine-sheath configuration located near the VLBI radio core (several parsecs away from the black hole), where photons of the slow sheath act as target photons for the faster spine.

The first four years of \fermi\ data on PKS\,1510-089 were analyzed in \cite{b13}. Hour-scale variability in the HE \g-ray band was established. Furthermore, the work tried to find the location of the emission region of the \g-rays within the jet. While no particular constraint was possible, the conclusion was drawn that the \g-rays cannot originate solely from the base of the jet, but probably come from several regions located throughout the jet.

VHE \g-ray variability in PKS\,1510-089 was finally established during another strong MWL flare in 2015 \cite{aMea17}. Similar models as during the 2012 flare \cite{aMea14} were used giving comparable results. As in both cases a connection with the emergence of a new moving radio knot on parsec scales is possible, the flares might be induced at that distance within the radio core. This would be far away from the usual positions within the thermal fields of the BLR or the DT.

As if the flares mentioned above were not enough, PKS\,1510-089 still had a surprise for observers. In May/June 2016 it underwent another major flare, which lasted merely two days and was concentrated on the highest energies \cite{zea18}. The VHE \g-ray flux rose by more than a factor 10 about previously measured values, while the integrated HE \g-ray flux barely changed. However, the HE \g-ray spectrum significantly hardened up to photon indices of $1.6$. The observations with \hess\ and MAGIC revealed for the first time intra-night variability in the VHE \g-ray band in this source. The only other contemporaneous observations were conducted with the optical telescope ATOM in the R-band. A mild flare can be seen in the optical lightcurve, however the details differ considerably from the VHE \g-ray lightcurve.

In the meantime, the MAGIC Collaboration integrated their data taken during HE \g-ray low states \cite{aMFea18}. These observations reveal a significant VHE \g-ray flux implying that for the first time an FSRQ has been established as a continuous VHE \g-ray emitter. This has strong consequences for the emission region and the emission processes at work within the jet.
%
%
\section{CTA\,102}
The last source to be discussed here is CTA\,102, an FSRQ at $z=1.032$ that was one of the first quasars to be detected in the mid 1960s\footnote{Fun fact: It even got a song by ``The Byrds'' in 1967.}. While it was known to be a variable source, during the first years of operation of \fermi, it was remarkably silent, and was barely detectable in daily bins. This changed in 2012, when the source flux increased significantly. With the exception of a few months in 2014, the source has been detectable on daily scales with \fermi\ ever since. The source behavior changed again in the beginning of 2016, when a strong flare raised the average flux level further. From November 2016 till February 2017, CTA\,102 underwent a spectacular outburst. Another outburst at the beginning of 2018 marked another change in the average flux back to pre-2016 levels \cite{zea19}.

This brief description of the last 10 years in HE \g-rays already shows what a remarkable source CTA\,102 is. In 2012 a bright MWL outburst was detected \cite{lea16}. It was found that the flux relationship between the optical and \g-ray flux was linear during the flare, but quadratic after the flare. The former has been interpreted as a change in the Doppler factor, while the latter is consistent with an SSC dominated \g-ray domain. Spectral shifts during the flare, as well as the Stokes parameter pattern support the conclusion of a varying Doppler factor as the main driver of the flare.

The flare in late 2016 and early 2017 was exceptionally bright. Peak fluxes in the \g-ray domain rose by more than a factor 50 over the pre-flare level (even though that was already a factor of a few above the pre-2012 quiescent flux), while the optical flux rose by a factor 100 over the pre-flare level\footnote{The optical flux became so bright that CTA\,102 was visible by eye through a small telescope.} \cite{zea17}. Fast variability was observed in both bands \cite{bea17,sea18}. A varying Doppler factor was again proposed as an explanation \cite{Rea17}, which include different (i.e. non-co-spatial) emission locations for the opical, infrared and radio emission, as these energy bands required different Doppler factors to explain the observed variability. A wobbling jet could explain the different Doppler factors exhibited by CTA\,102 during the flare and the previous years.

Alternatively, the big flare could have been induced by the ablation of a gas cloud \cite{zea17,zea19} that intruded into the jet. This process continuously injects matter from the cloud into the jet flow. The duration of the event is then determined by the size of the cloud, and the flux enhancement by the density of the cloud. The model provides a relatively simple explanation for the event with minimal assumptions that can easily explain the symmetry of the lightcurve of the 4-months-long flare. An open question in this model is the nature of the gas cloud. Possibilities with different probabilities range from BLR clouds over star-forming regions to the astrospheres of red giant stars.

VLBI radio observations \cite{cea19} around the flare have revealed the emergence of a superluminal radio knot with an estimated variability Doppler factor of $\sim 35$ before the flare, which crossed a recollimation shock during the time of the flare. This supports both models described above.
%
%
\section{Summary}
%
In this paper a brief overview was given on surprising blazar activity. The four listed sources -- 3C\,279, PKS\,2155-304, PKS\,1510-089, and CTA\,102 -- exhibited vastly different flares in terms of duration, lightcurve profile, spectral behavior and MWL correlations. This is true not just from source to source but also for the sources individually, as no flare has been comparable to another.
The typical (leptonic) one-zone model faces severe difficulties in explaining any of these flares, and many other models have been invoked with varying successes. The VHE \g-ray detections of FSRQs complicate the picture substantially. They imply that the responsible emission region cannot be located deep within the BLR \cite{cea18}, as the inevitable absorption would not allow the detection of VHE \g-rays. Hence, the emission regions are located relatively far away from the black hole, where the jet may have widened considerably already. This implies that the emission regions of the fast flares do not encompass the entire width of the jet, and point instead to small subregions within a larger flaring environment \cite{g13,m14}.

Up to this point an ``encompassing picture'' of blazar flares remains elusive. Especially, the missing similarity of flares within a single source complicates the picture, as these flares should originate from within the same large-scale jet. It will be an interesting task to ``evolve'' the different types of flares from -- more or less -- the same ground state. For now, this makes it impossible to predict how a flare evolves even on short time-scales. On the other hand, it encourages deep observations at all wavelengths -- at best, in the form of monitoring -- in order to obtain the complete picture for the whole duration of a flare.

%
%
\section*{Acknowledgement}
I wish to thank the organizers of the HEASA\,2018 conference for the invitation and their hospitality. The list of sources and referenced flares and models is a personal choice, and certainly incomplete. My sincere apologies to authors of omitted papers. \\
Funding by the German Ministry for Education and Research (BMBF) through grant 05A17PC3 is gratefully acknowledged.
%
%

%
\end{document}